\documentclass[aps,prd,twocolumn,nopacs,floatfix,amsmath,nofootinbib,amssymb,preprintnumbers]{revtex4-2}
\usepackage{braket}
\usepackage{graphicx,color,dcolumn,booktabs}
\usepackage{array}
\usepackage{tabularx}
\usepackage{txfonts}
\usepackage{slashed}
\usepackage{epstopdf}
\usepackage{amsmath}
\usepackage{appendix}
\usepackage{multirow} % Add to your preamble
\usepackage[table]{xcolor} % Already included if using \rowcolor
\usepackage[colorlinks,
citecolor=blue,
anchorcolor=red,
menucolor=red,
linkcolor=red,
filecolor=red,
runcolor=red,
urlcolor=blue,
frenchlinks=true]{hyperref}
\usepackage{bm}

\graphicspath{{Figures/}}
\begin{document}
%Unquenching effects on $P$-wave charmonia $\chi_{cJ}$ in nuclear matter
\title{Medium modifications of $1P$-wave charmonia $\chi_{cJ}(1P)$ in cold nuclear matter}
\author{Ze-Hua Zhang$^{1,2,4}$}\email{220220940191@lzu.edu.cn}
\author{Xiang Liu$^{1,2,3,4}$}\email{xiangliu@lzu.edu.cn}

\affiliation{
$^1$School of Physical Science and Technology, Lanzhou University, Lanzhou 730000, China\\
$^2$Lanzhou Center for Theoretical Physics,
        Key Laboratory of Theoretical Physics of Gansu Province,
        Key Laboratory of Quantum Theory and Applications of MoE,
        Gansu Provincial Research Center for Basic Disciplines of Quantum Physics, Lanzhou University, Lanzhou 730000, China\\
        $^3$MoE Frontiers Science Center for Rare Isotopes, Lanzhou University, Lanzhou 730000, China\\
        $^4$Research Center for Hadron and CSR Physics, Lanzhou University $\&$ Institute of Modern Physics of CAS, Lanzhou 730000, China}

\begin{abstract}
In this work, we employ the quark-meson coupling model to investigate the mass shifts of $1P$-wave charmonia $\chi_{cJ}(1P)$ ($J=0,1,2$) in cold symmetric nuclear matter by incorporating in-medium loop contributions to the $\chi_{cJ}(1P)$ self-energy within the unquenched picture. At normal nuclear matter density, we obtain significant mass reductions of about 60 MeV for the $\chi_{cJ}(1P)$ states, with the $\chi_{c2}(1P)$ mass shift primarily arising from the vector–vector loop. Our results also indicate the absence of level crossing between the in-medium $\chi_{cJ}(1P)$ mass and the $D\bar{D}$ mass threshold up to $\rho_B < 3\rho_0$—a feature that could be probed in the Compressed Baryonic Matter experiment at FAIR and the Beam Energy Scan program at RHIC.
\end{abstract}

\maketitle

\section{INTRODUCTION}

The study of heavy quarkonia provides a unique probe into the non-perturbative behavior of the strong interaction. In particular, while the vacuum charmonium spectrum is well understood through quenched potential models~\cite{PhysRevD.21.203,Godfrey:1985xj,Voloshin:2007dx,Brambilla:2010cs}\textemdash with unquenching effects included as essential corrections~\cite{Kalashnikova:2005ui,Chao:2007it,Li:2009zu,Duan:2020tsx}\textemdash the behavior of charmonia at finite density and/or temperature has only recently garnered increased attention~\cite{Saito:2005rv,Brambilla:2010cs,Hosaka:2016ypm,Krein:2017usp,Rothkopf:2019ipj}. Possible medium-induced modifications of charmonia have been explored in various contexts, including $(c\bar{c})$-nucleus bound states~\cite{Brodsky:1989jd,Wasson:1991fb,Kaidalov:1992hd,Belyaev:2006vn,Tsushima:2011kh,Yokota:2013sfa,Yokota:2014lma}, the behavior of charmonium-like states such as $X(3872)$ under extreme conditions~\cite{ExHIC:2010gcb,ExHIC:2011say,Cho:2013rpa,Abreu:2016qci,ExHIC:2017smd,Azizi:2017ubq,Zhang:2020dwn,Albaladejo:2021cxj}, in-medium spectral properties of charmonia~\cite{7zby-gy46,Rapp:2008tf}, and the suppression of $J/\psi$ production as evidence of quark-gluon plasma formation~\cite{Matsui:1986dk}.

Among various theoretical approaches to studying charmed hadrons at finite density, the quark-meson coupling (QMC) model has emerged as an effective framework for describing hadronic matter in nuclear matter, finite nuclei, and neutron stars~\cite{GUICHON1988235,GUICHON1996349,Rikovska-Stone:2006gml}. Pioneering in treating hadrons as extended objects, the model attributes the binding of nucleons in nuclear matter to self-consistent couplings between confined light quarks inside nucleons and the surrounding mesonic mean fields. The treatment of heavy quarkonia in this model is distinctive, as their constituent quarks differ from those in the surrounding nuclear medium. Specifically, medium modifications of charmonia such as $J/\psi$ are incorporated via excitations of $D^{(*)}\bar{D}^{(*)}$ meson loops, which contain light quarks~\cite{Krein:2010vp}. In fact, it is a typical unquenched picture, which has been extensively applied to charmonium spectroscopy in free space, particularly for high-lying states~\cite{Wang:2019mhs,Wang:2020prx}. It has also helped resolve low-mass puzzles of certain charmed hadrons~\cite{vanBeveren:2003kd,Dai:2003yg,Hwang:2004cd,Simonov:2004ar,Danilkin:2010cc,Li:2009ad,Kalashnikova:2005ui,Luo:2019qkm}, and has inspired novel mechanisms such as the formation of $X(6900)$ via di-$J/\psi$ interactions~\cite{Huang:2024jin}, as well as predictions of bound states between conventional hadrons and hadronic molecules, such as $P_c N$ systems~\cite{Qian:2024joy}. This framework also describes medium modifications of other heavy quarkonia and two-flavored heavy mesons like $\eta_b$, $\Upsilon$, and $B_c$, enabling the evaluation of their bound-state energies with finite nuclei within the QMC model~\cite{Tsushima:2011kh,Cobos-Martinez:2020ynh,Zeminiani:2020aho,Cobos-Martinez:2022fmt,Zeminiani:2024dyo,Zeminiani:2023gqc}.

In these studies, as well as in updated predictions for $J/\psi$ mass shifts~\cite{Krein:2017usp,Zeminiani:2020aho}, contributions from the vector-vector channel were acknowledged but ultimately omitted due to their relatively large impact on the predicted mass shifts. In particular, the sizable effect from the heavier $D^{*}\bar{D}^{*}$ loop was considered unexpected compared to that of the $D\bar{D}$ loop. Consequently, only the so-called “lowest-order meson loops,” typically restricted to vector-pseudoscalar or pseudoscalar-pseudoscalar configurations, were retained, and the cutoff mass was readjusted. Although predictions for bound-state formation may remain stable over a broad range of cutoff parameters, the vector-vector channel could play a crucial role in the in-medium mass modification of other heavy quarkonia. For this theoretical purpose, $\chi_{c2}(1P)$ is particularly relevant, as its $D$-wave coupling to $D\bar{D}$, $D^{*}\bar{D}$, or $D\bar{D}^{*}$ is suppressed compared to its $S$-wave coupling to the $D^{*}\bar{D}^{*}$ loop.

Phenomenologically, modifications of the $\chi_{cJ}(1P)$ triplet under temperature and density are critical for understanding the observed suppression of $J/\psi$ production, given that approximately $40\%$ of observed $J/\psi$ events originate from feed-down processes involving $\chi_{c1}(1P)$, $\chi_{c2}(1P)$ and $\psi(2S)$ states~\cite{E705:1992jno,E705:1992vec}. $J/\psi$ suppression has been widely reported in nucleus-nucleus collisions~\cite{NA50:1996lag,NA50:1997hlx,NA50:1999stx,NA50:2000brc,PHENIX:2006gsi,PHENIX:2011img,CMS:2012bms,ALICE:2012jsl}, as well as in proton (deuteron)-nucleus collisions~\cite{NuSea:1999mrl,Scomparin:2009tg,ALICE:2013snh}. Thus, investigations of various cold nuclear matter (CNM) effects, induced by the presence of nuclei in the initial state, serve as a baseline for understanding quarkonia production and suppression~\cite{Albacete:2013ei,Albacete:2017qng,Andronic:2015wma}. Analogous to the sequential dissociation of charmonia in hot media~\cite{Karsch:2005nk}, a similar pattern may arise with increasing nuclear density\textemdash known as the level crossing effect~\cite{Hayashigaki:2000es,Mishra:2003se,Friman:2002fs}. It has been debated whether the $D\bar{D}$ mass threshold could sequentially drop below the masses of excited charmonia $\psi(2S)$, $\chi_{c2}(1P)$, $\chi_{c1}(1P)$, and $\chi_{c0}(1P)$, potentially leading to a stepwise suppression of $J/\psi$ in a hadronic dissociation scenario~\cite{Sibirtsev:1999jr,Golubeva:2002jz}.

Various approaches have been employed to estimate charmonium mass shifts in nuclear matter, including chiral effective models~\cite{Kumar:2010gb,Kumar:2011ff}, QCD sum rules~\cite{Klingl:1998sr,Hayashigaki:1998ey,Kim:2000kj,Kumar:2010hs}, and other methods~\cite{Luke:1992tm,Lee:2000csl,Sibirtsev:2005ex,TarrusCastella:2018php}. Although QCD sum rules have been used to estimate in-medium masses of $\chi_{cJ}(1P)$ mesons~\cite{Lee:2003hm,Morita:2007hv,Song:2008bd,Morita:2010pd,Kumar:2018cwk}, these studies generally predict nearly degenerate mass shifts of about $\Delta M_{\chi_{cJ}} \approx -60~\mathrm{MeV}$ for $J=0,1,2$ at zero temperature and normal nuclear density $\rho_0$~\cite{Lee:2003hm,Morita:2007hv}. However, no detailed analysis of the $\chi_{cJ}(1P)$ system has been performed within the QMC model, as the vector-vector channel was omitted in earlier treatments.

In this work, we investigate the in-medium mass of the $1P$-wave charmonia $\chi_{cJ}(1P)$ in cold symmetric nuclear matter using the QMC model and heavy quark effective field theory. By systematically evaluating loop contributions from different intermediate states, we highlight the non-negligible role of the vector-vector loop in the $\chi_{cJ}(1P)$ mass shift, especially for $\chi_{c2}(1P)$. This finding implies that although the $D\bar{D}$ mass threshold decreases with increasing density, the effective masses of the $\chi_{cJ}(1P)$ triplet remain below this threshold, thereby preventing a stepwise $J/\psi$ suppression signal that might be expected if $1P$-wave charmonia experienced mass shifts similar to $J/\psi$. Furthermore, the obtained in-medium masses provide a valuable theoretical input for modeling various CNM effects.

This paper is organized as follows. In Sec.~\ref{sec:model}, we present the essential elements of the QMC model and its predictions for the in-medium mass of $D^{(*)}$ mesons, along with the effective Lagrangian used to compute the $\chi_{cJ}(1P)$ self-energy and the explicit expressions for different intermediate states. In Sec.~\ref{sec:results}, after determining the relevant parameters, we present our numerical results for the mass shifts of $\chi_{c0}(1P)$, $\chi_{c1}(1P)$, and $\chi_{c2}(1P)$ mesons and discuss notable features. Finally, a summary and future perspectives are provided in Sec.~\ref{sec:summary}.

\section{Theoretical Framework}
\label{sec:model}

At the hadronic level, the interaction between charmonium and nucleons in nuclei is suppressed due to the Okubo–Zweig–Iizuka (OZI) rule, which indicates that heavy quarks do not couple directly to light mesons. Thus, in-medium modifications of charmed mesons play a crucial role in the one-loop contribution considered in this work, with their Lorentz-scalar effective masses evaluated within the QMC model. In this section, we first review the essential aspects of this model and then introduce how medium modification of intermediate states such as $D^{(*)}$ and $\bar{D}^{(*)}$ correct the mass spectrum of $P$-wave charmonia. 

The QMC model, which provides a natural mechanism for nuclear saturation, was originally proposed by Guichon in Ref.~\cite{GUICHON1988235} to describe nuclear matter and later applied to finite nuclei in Ref.~\cite{GUICHON1996349}. This quark-based model emphasizes  changes in the internal structure of nucleons in a finite-density environment, implemented using the MIT bag model~\cite{Chodos:1974je,DeGrand:1975cf} and assuming that the scalar-$\sigma$ and vector-$\omega$ meson mean fields couple directly to the light (anti)quarks inside non-overlapping nucleon bags. For further details on the QMC model, including predictions and applications, see Refs.~\cite{Saito:2005rv,Krein:2017usp,Guichon:2018uew}.

In this study, we restrict our analysis to the cold symmetric nuclear matter. The Lagrangian density depicting this infinite many-nucleon system in the QMC model is given by~\cite{GUICHON1996349}
\begin{eqnarray}\label{eq:0-orderQMC}
     \mathcal{L}^{0}_{\text{QMC}}&=&\bar{\psi}_{N}\big[i \gamma \cdot \partial - M^{*}_{N}\big(\hat{\sigma}\big)-g_{\omega} \hat{\omega}^{\mu}\gamma_{_\mu} \big]\psi_{N}\notag\\
     &&+\frac{1}{2}\big(\partial_{\mu} \hat{\sigma}\partial^{\mu} \hat{\sigma} -m^{2}_{\sigma}{\hat{\sigma}}^2\big)\notag\\
     &&-\frac{1}{2}\big[ \partial_{\mu} \hat{\omega}_{\nu}\big(\partial^{\mu} \hat{\omega}^{\nu}-\partial^{\nu} \hat{\omega}^{\mu}\big) -m_{\omega}^{2} \hat{\omega}^{\mu}\hat{\omega}_{\mu} \big],
\end{eqnarray}
where $\psi_N$, $\hat{\sigma}$ and $\hat{\omega}^{\mu}$ denote the nucleon, attractive $\hat{\sigma}$ field, and repulsive $\hat{\omega}^{\mu}$ field, respectively. Here, $m_\sigma$ and $m_\omega$ are the masses of the scalar and vector meson fields, and $g_{\omega}$ is the coupling constant for the $\omega$-$N$ interaction at hadronic level. Equation~(\ref{eq:0-orderQMC}) is derived under the mean field approximation, where the operators $\hat{\sigma}$ and $\hat{\omega}^{\mu}$ are treated as classical fields. Using the translation and rotation symmetry of nuclear matter, the following replacements are made in the nuclear matter rest frame
\begin{eqnarray}
     \hat{\sigma} \to \braket{\hat{\sigma}} \equiv \sigma, \quad \hat{\omega}^{\mu} \to \braket{\hat{\omega}^{\mu}} \equiv {\delta}^{\mu 0} \omega,
\end{eqnarray}
where the expectation value $\sigma$ and $\omega$ are taken with respect to a uniform bulk of nuclear matter. This approach can be generalized to finite nuclei using  the local density approximation~\cite{Tsushima:1998ru}.

The superscript in $\mathcal{L}^{0}_{\text{QMC}}$ indicates that $\rho$ meson contribution and Coulomb interaction are not included here; these are treated as corrections in studies of finite nuclei~\cite{GUICHON1996349,Tsushima:1998ru,Saito:1996yb,Cobos-Martinez:2023hbp,Zeminiani:2024dyo} and systems with asymmetric proton-neutron distributions~\cite{Mondal:2023iwe,Mondal:2024vyt,Mondal:2025qxm}. The effective nucleon mass $M^{*}_N$ is defined as
\begin{eqnarray}\label{eq:hlevelmass}
     M^{*}_{N}\big(\sigma\big) \equiv M_{N}-g_{\sigma}\big(\sigma\big) \cdot \sigma,
\end{eqnarray}
where the field-dependent coupling constant $g_{\sigma}\big(\sigma\big)$ incorporates the internal structure of nucleons. In the QMC model, this structure is described using the MIT bag model.

Assuming confinement via a linear boundary condition and introducing a bag constant 
$B$ to balance the outward pressure from quark and antiquark motion, the Dirac equations for quarks inside the bag are given by~\cite{Tsushima:1997df,Tsushima:2002cc}
\begin{eqnarray}\label{eq:qleveldiraceq}
\Big[i\gamma \cdot \partial -\big(m_{q} - V^{q}_{\sigma}\big) \mp \gamma^0 V^{q}_{\omega}\Big] \begin{pmatrix} \psi_{q}  
\\ \psi_{\bar{q}} \end{pmatrix} &=& 0,\\
\Big[i\gamma \cdot \partial - m_Q \Big] \psi_{Q,\bar{Q}}&=& 0,
\end{eqnarray}
where $m_{q}$ denotes the light quark mass $(q=u,d)$, and heavy quarks $(Q=s,c,b)$ are assumed not to couple directly to the mean fields. The mean-field potential in Eq.~(\ref{eq:qleveldiraceq}) takes the form $V^{q}_{\sigma}=g^{q}_{\sigma}\sigma$ and $V^{q}_{\omega}=g^{q}_{\omega}\omega$, with $g^{q}_{\sigma}$ and $g^{q}_{\omega}$ being the quark-level coupling constants.

The ground-state energy of an in-medium (anti)quark in a hadron can be expressed as a function of $\sigma$, $\omega$ and bag radius $R^{*}_{h}$:
\begin{eqnarray}
\left( \begin{array}{c} \epsilon^*_{q} \\ \epsilon^*_{\bar{q}} \end{array} \right)
&=& \Omega_{q}^* \pm R^*_{h} V^{q}_\omega ,\\ \label{B27}
\epsilon^*_{Q} &=& \epsilon_{\bar{Q}}^*= \Omega^*_{Q},\\\notag
\end{eqnarray} 
where $h$ denotes hadrons such as $D^{(*)}$ or $\bar{D}^{(*)}$. For a (anti)quark of flavor $(f=q,Q,\bar{q},\bar{Q})$, the symbol $x^{*}_{f}$ denotes the lowest-mode bag eigenfrequency, and $\Omega^*_{f}\big(\sigma,R^{*}_{h}\big)=\big[x^{*2}_{f}+\big(R^{*}_{h}m_{f}^{*}\big)^2\big]^{\frac{1}{2}}$, where $m^{*}_{Q}=m^{*}_{\bar{Q}}=m_{Q}$ and $m^{*}_{q}=m^{*}_{\bar{q}}=m_{q}-g^{q}_{\sigma}\sigma$. Using the linear boundary condition, $x^{*}_{f}\big(\sigma,R^{*}_{h}\big)$ is determined via
\begin{eqnarray}\label{eq:qmcmass}
     j_{0}\big(x_{f}^{*}\big)=\sqrt{\frac{\Omega^*_{f}-m_{f}^{*}R^{*}_{h}}{\Omega^*_{f}+m_{f}^{*}R^{*}_{h}}}j_{1}\big(x_{f}^{*}\big), 
\end{eqnarray}
where $j_{0}$ and $j_{1}$ are spherical Bessel functions.

The effective mass of a hadron bag in medium is then given by
\begin{eqnarray}
     &&m_{h}^{*}\big(\sigma\big)=E^{\text{bag}}_{h}\big(\sigma\big)=\sum_{f} \frac{n_{f}\Omega^{*}_{f}-z_{h}}{R_{h}^{*}}+\frac{4}{3} \pi R_{h}^{* 3} B,\label{eq:hadronmass}\\
     &&\left. \frac{\partial m^{*}_{h}\big(\sigma,R_{h}^{'}\big)}{\partial R_{h}^{'}}\right| _{R_{h}^{'}=R_{h}^{*}}=0,\label{eq:nonlinearbc}
\end{eqnarray}
where $n_{f}$ is the number of quarks of given flavor in the hadron bag, and $z$ parametrizes center-of-mass and gluon fluctuation effects~\cite{GUICHON1996349}. The bag parameters $B$ and $z_{h}$ are fixed by fitting the proton and charmed meson masses in vacuum using Eqs.~(\ref{eq:hadronmass}) and (\ref{eq:nonlinearbc}). The nonlinear boundary condition Eq.~(\ref{eq:nonlinearbc}) determines the in-medium bag radius $R_{h}^{*}\big(\sigma\big)$, which varies with the $\sigma$ mean field and depends on the specific hadron $h$.

We note that mass splitting between $D$ and $D^{*}$ mesons in the MIT bag model~\cite{Shuryak:1980pg,Izatt:1981pt,Sadzikowski:1993uv,Sadzikowski:2003jy} can in principle be described by introducing hyperfine structure and evaluating the $D^{*}$ in-medium mass with an additional term in Eq.~(\ref{eq:qmcmass}). However, since the unquenched effect on charmonia (such as $\chi_{cJ}$ in this work) arises from the loop diagram, we neglected this higher-order contribution in the present work. Instead, mass splittings due to spin-spin interactions were included by assigning different bag radii ($R_{D} \neq R_{D}^{*}$) and parameters ($z_{D} \neq z_{D^{*}}$), following a  treatment similar to that in Ref.~\cite{Tsushima:2002cc}. The same approach was applied to $\bar{D}$ and $\bar{D}^{*}$.

The relationship between the quark-level coupling $g_{\sigma}^{q}$ and the field-dependent hadronic coupling $g_{\sigma}(\sigma)$, as well as the corresponding relation for the $\omega$ mean-field coupling, is given by
\begin{eqnarray}
     \partial_{\sigma}\big[g_{\sigma}\big(\sigma\big) \cdot \sigma\big]&=&3 g^{q}_{\sigma} S\big(\sigma\big),\\
     g_{\omega}&=&3 g_{\omega}^{q},
\end{eqnarray}
where
\begin{eqnarray}
     S\big(\sigma\big)=\frac{\Omega_{q}^*/2+m_{q}^{*} R_{N}^{*} (\Omega_{q}^{*}-1)}{\Omega_{q}^{*}(\Omega_{q}^{*}-1)+m_{q}^{*} R_{N}^{*}/2}
\end{eqnarray}
encodes the response of the nucleon structure (in this case, the MIT bag) to the external scalar field $\sigma$. These relations are derived by comparing equations of motion at the quark-level with those from Eq.~(\ref{eq:0-orderQMC}). As a detailed analysis of the self-consistency equation relating the $\sigma$ mean field to the baryon density $\rho_{B}$, see Ref.~\cite{GUICHON1996349}.

Based on the above formalism, we present QMC predictions for the in-medium mass of $D^{(*)}$ mesons in symmetric nuclear matter. The parameters are fixed as follows: current quark masses $m_{q}=m_{u}=m_{d}=5~\text{MeV}$ and $m_{c}=1300~\text{MeV}$, a free nucleon mass $ m_N = 939.0~\text{MeV}$, and a free-space nucleon bag radius $R_{N}=0.8~\text{fm}$, consistent with standard QMC input. Using Eqs.~(\ref{eq:hadronmass}) and (\ref{eq:nonlinearbc}), the nucleon bag parameters are determined as \( B = (170~\text{MeV})^4 \) and \( z_N = 3.295 \). Similarly, we adopt $m_D=1867.2~\text{MeV}$ ( average of the neutral and charged $D$ masses) and $m_{D^{*}}=2008.6~\text{MeV}$ (average of $D^{*+}$ and ${D^{*0}}$ masses) to determine bag radii and  parameters for $D^{*}$ and $\bar{D}^{*}$ mesons, as listed in Table~\ref{tab:bagparam}. Standard meson masses $m_{\sigma}=550~\text{MeV}$ and $m_\omega=783~\text{MeV}$ are used, along with light quark-meson coupling constants $g^{q}_{\sigma}=\frac{1}{3} g_{\sigma}\big(0\big) S\big(0\big)=5.69$ and $g^{q}_{\omega}=\frac{1}{3} g_{\omega}=2.72$. These values are consistent with Refs.~\cite{Tsushima:2011kh,Saito:2005rv}, where they were calibrated to reproduce the saturation properties and symmetry energy of nuclear matter.

\begin{table}[htbp]
\begin{center}
\caption{
Physical masses in free space (used as input), along with corresponding bag parameters $z$ and bag radii $R$. In-medium effective masses $m^{*}$ are calculated at normal nuclear matter density $\rho_{0}=0.15~\text{fm}^{-3}$ using the QMC model with the bag constant $B=(170.0~\text{MeV})^{4}$ and current quark masses $m_{u}=m_{d}=5~\text{MeV}$ and $m_{c}=1300~\text{MeV}$.}
\label{tab:bagparam}
\renewcommand\tabcolsep{0.23cm}
\renewcommand{\arraystretch}{1.5}
\begin{tabular}[t]{ccccc}
\toprule
\toprule
Hadron &Mass (MeV) &$z$ &$R$ (fm)& $m^{*}$ (MeV) \\
\hline
$N$            &939.0 (input) &3.295 &0.800 (input) &753.5  \\
$D,\bar{D}$   &1867.2 (input)&1.388 &0.731         &1804.8 \\
$D^{*},\bar{D}^{*}$  &2008.6 (input)&0.849 &0.774 &1946.5 \\
\bottomrule
\bottomrule
\end{tabular}
\end{center}
\end{table}

At normal nuclear matter density $\rho_{B}=\rho_{0}$, the mass shift of the $D^*$ meson is approximately $-62.1$~MeV, and that of the $D$ meson is about $-62.4$~MeV. The parameters and mass shift we obtain for charmed mesons are consistent with Ref.~\cite{Tsushima:1998ru}.

The absence of light quark content in $1P$-wave charmonia $\chi_{cJ}(1P)$ implies that, within the QMC model, they do not couple directly to nuclear matter. Instead, their in-medium modifications are incorporated through an unquenched picture, which has been well established for high-lying charmonia as coupled-channel effects. In the present work, an effective Lagrangian approach is used to describe the loop contribution to the $\chi_{cJ}(1P)$ self-energy~\cite{Wise:1992hn,Burdman:1992gh,Casalbuoni:1996pg,Yan:1992gz}. The explicit form of the interaction Lagrangian is given by~\cite{Huang:2021kfm,Colangelo:2003sa,Chen:2013cpa,Chen:2014ccr}:
\begin{eqnarray}\label{eqs:chicjDD}
&&\mathcal{L}_{\chi_{cJ} D^{(\ast)} D^{(\ast)}}\notag\\
&&\quad=- g_{\chi_{c0} D D } \chi_{c0} \bar{D}
D - g_{\chi_{c0} D^\ast
D^\ast} \chi_{c0} \bar{D}_{\mu}^\ast D^{\ast
\mu} \notag\\
&&\quad\quad +i g_{\chi_{c1} D
D^\ast} \chi_{c1}^\mu ( \bar{D} D^{\ast }_\mu - \bar{D}^{\ast}_\mu D  )
\notag\\
&&\quad\quad - g_{\chi_{c2} D D} \chi_{c2}^{\mu \nu}
\partial_\nu \bar{D} \partial_\mu D + g_{\chi_{c2} D^\ast D^\ast} \chi_{c2}^{\mu \nu} \bar{D}^\ast_{\mu} D^{\ast}_\nu \notag\\
&&\quad\quad -ig_{\chi_{c2} D^\ast D} \varepsilon_{\mu
\nu \alpha \beta} \partial^\alpha \chi_{c2}^{\mu \rho}
(\partial^\beta \bar{D} \partial_\rho D^{\ast \nu} -\partial_\rho \bar{D}^{\ast \nu} \partial^\beta D),
\end{eqnarray}
where we adopt the convention $D^{(\ast)}=\big(D^{(\ast)+}\,, D^{(\ast)0}\big)^{\text{T}}$ and $\bar{D}^{(\ast)}=\big(D^{(\ast)-}\,, \bar{D}^{(\ast)0}\big)$. Contributions from $D_s^{(*)+}$ mesons, which have $c\bar{s}$ quark content, are not considered here~\cite{Krein:2010vp}; the same applies to $D_s^{(*)-}$ meson.

Denoting the in-medium mass of $1P$-wave charmonia by $M^{*}_{\chi_{cJ}}$, their modification in nuclear matter is described by the density-dependent relation:
\begin{eqnarray}\label{eq:massinmedium}
     ({M^{*}_{\chi_{cJ}}})^2 =\big(M_{\chi_{cJ}}^{0}\big)^{2} + \sum_{l} \Re \Pi_{l}^{*}\big[({M^{*}_{\chi_{cJ}}})^2\big],
\end{eqnarray}
where $\Pi_{l}^{*}$ is the scalar self-energy from unquenched hadronic loops at a given baryon density, with free-space mass $m_{D^{(*)}}$, $m_{\bar{D}^{(*)}}$ and $M_{\chi_{cJ}}$ replaced by their in-medium form counterparts.

The scalar self-energy from the loops, derived from the interactions in Eq.~(\ref{eqs:chicjDD}), takes the generic form:
\begin{eqnarray}\label{eq:scalarselfenergy}
\Pi_{l}(p^2)=\int\frac{d^{4}q}{(2\pi)^{4}}	\frac{{\mathcal V}_{1}{\mathcal V}_{2}}{{\mathcal P}_{1}{\mathcal P}_{2}}\mathcal F_{1}(q^{2},m_{D^{(*)}}^2)\mathcal F_{2}((p-q)^{2},m_{\bar{D}^{(*)}}^2),\notag\\
\end{eqnarray}
where $l$ labels different loops: $D\bar{D}$, $D^*\bar{D}$, $D\bar{D}^*$ or $D^*\bar{D}^*$. Here, $1/\mathcal{P}_i$ ($i=1,\,2$) represent propagators of intermediate $D^{(*)}$ and $\bar{D}^{(*)}$ mesons, while $\mathcal{V}_{i}$ comes from the interaction vertices. A dipole form factor $\mathcal{F}_{i}(q^2,m_E^2)=\left(\frac{m_E^2-\Lambda^2}{q^2-\Lambda^2}\right)^2$ is introduced to regularize the integral and ensure amplitude convergence, where $m_E$, $q$ and $\Lambda$ are the mass, four-momentum of the exchanged meson, and cutoff momentum, respectively.

\begin{figure}[htbp]
    \centering
    \includegraphics[width=0.48\textwidth]{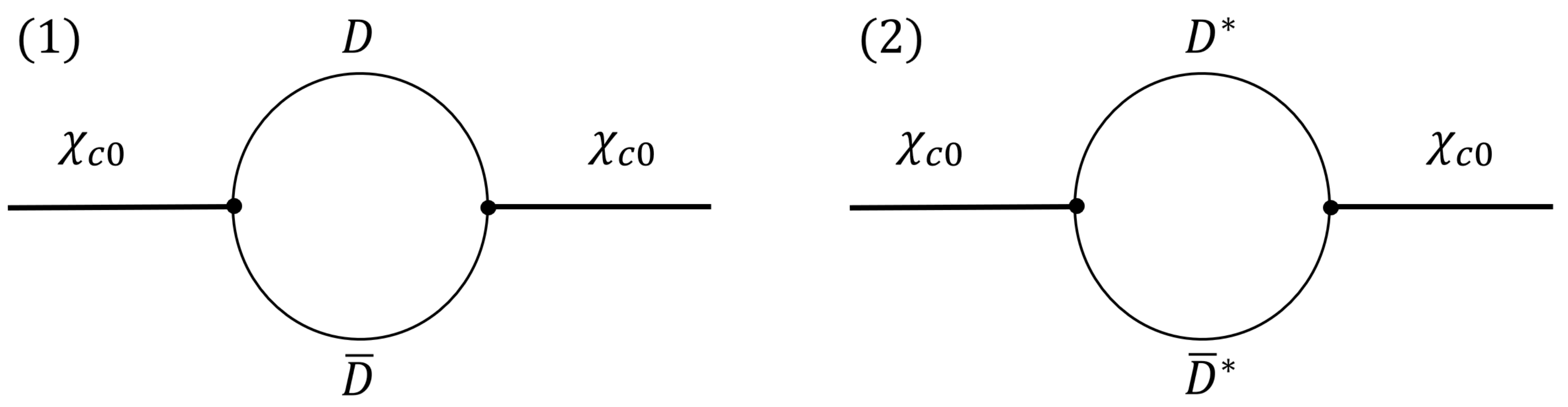}
    \caption{Schematic diagram of the charmed-meson loop contributions to the $\chi_{c0}(1P)$ self-energy, where (1) and (2) represent the $D\bar{D}$ and $D^{*}\bar{D}^{*}$ loop contributions, respectively.}
    \label{fig:chic0loop}
\end{figure}

According to the effective Lagrangian, the Feynman diagram for $\chi_{c0}(1P)$ is shown in Fig.~\ref{fig:chic0loop}. The $D \bar{D}$ and $D^{*} \bar{D}^{*}$ loop contributions to the $\chi_{c0}(1P)$ self-energy are given by
\begin{eqnarray}
     \Pi^{\text{total}}_{\chi_{c0}}=\Pi_{\chi_{c0}D\bar{D}}+\Pi_{\chi_{c0}D^{*}\bar{D}^{*}},
\label{eq:chic0totloop}
\end{eqnarray}
where
\begin{eqnarray}
\Pi_{\chi_{c0}D\bar{D}}(p^2)&=&2ig_{\chi_{c0}D\bar{D}}^2  \int\frac{d^4 q}{(2 \pi)^4} \frac{1}{(q^2-m_{D}^2)((p-q)^2-m_{D}^2)}\notag\\
&&\times \mathcal{F}_{D}\big(q^2,m_{D}^{2}\big) \mathcal{F}_{D}\big((p-q)^2,m_{\bar{D}}^{2}\big),\label{eq:chic0dd}\\
\Pi_{\chi_{c0}D^{*}\bar{D}^{*}}(p^2)&=&2ig_{\chi_{c0}D^{*}\bar{D}^{*}}^2  \int\frac{d^4 q}{(2 \pi)^4} \frac{4 - \frac{q^2 + (p - q)^2}{m_{D^*}^2} + \frac{(q \cdot (p - q))^2}{m_{D^*}^4}}{(q^2-m_{D^*}^2)((p-q)^2-m_{D^*}^2)}\notag \\
&&\times  \mathcal{F}_{D}\big(q^2,m_{D^{*}}^{2}\big) \mathcal{F}_{D}\big((p-q)^2,m_{\bar{D}^{*}}^{2}\big).\label{eq:chic0dsds}
\end{eqnarray}

\begin{figure}[htbp]
    \centering
    \includegraphics[width=0.48\textwidth]{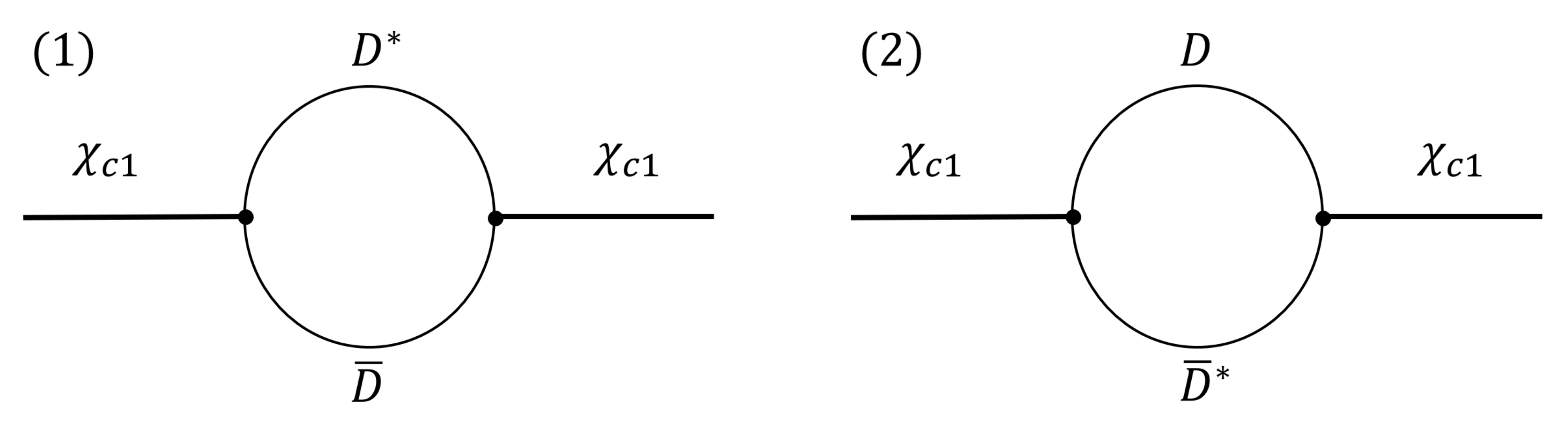}
    \caption{Schematic diagram of the charmed-meson loop contributions to the $\chi_{c1}(1P)$ self-energy, where (1) and (2) represent the $D^{*}\bar{D}$ and $D\bar{D}^{*}$ loop contributions, respectively.}
    \label{fig:chic1loop}
\end{figure}

Figure~\ref{fig:chic1loop} shows the loops considered for the $\chi_{c1}(1P)$ meson. To extract the scalar part of $\chi_{c1}(1P)$ self-energy that contributes directly to the physical mass, we use the projection operators $\tilde{g}_{\mu\nu}(p)=g_{\mu\nu}-\frac{p_{\mu}p_{\nu}}{p^2}$ and $\tilde{N}_{\mu\nu}(p)=\frac{p_{\mu}p_{\nu}}{p^2}$. The result is
\begin{eqnarray}
     \Pi^{\text{total}}_{\chi_{c1}}=\Pi_{\chi_{c1}D^*\bar{D}}+\Pi_{\chi_{c1}D\bar{D}^{*}}=2\Pi_{\chi_{c1}D^*\bar{D}},
\label{eq:chic1totloop}
\end{eqnarray}
where
\begin{eqnarray}
     \Pi_{\chi_{c1}D^{*}\bar{D}}(p^2)&=&-\frac{i}{3} \tilde{g}_{\mu\nu} i\Pi_{\chi_{c1}D^{*}\bar{D}}^{\mu\nu}\notag\\
        &=&\frac{2 i}{3} g_{\chi_{c1}D^{*}\bar{D}}^2 \int\frac{d^4 q}{(2 \pi)^4} \frac{4-\frac{q^2}{m_{D^*}^2}-1+\frac{(q\cdot p)^2}{p^2 m_{D^*}^2}}{(q^2-m_{D^*}^2)((p-q)^2-m_{\bar{D}}^2)}\notag\\
        &&\times\mathcal{F}_{D^{*}}\big(q^2,m_{D^{*}}^{2}\big)\mathcal{F}_{\bar{D}}\big((p-q)^2,m_{\bar{D}}^{2}\big).
        \label{eq:chic1oneloop}
\end{eqnarray}

For propagators with spin greater than spin-$\frac{3}{2}$, their explicit forms remain under discussion~\cite{PhysRevD.2.2255,PhysRev.173.1608,Ahluwalia:1991cf}. Our convention for the $\chi_{c2}(1P)$ meson follows Ref.~\cite{Huang:2005js}, though the projected loop contribution to the in-medium mass is convention-independent. We use the projection operators
\begin{eqnarray}\label{eq:projector2}
    P^{\mathrm{mass}}_{\mu \nu \rho \sigma}=\frac{1}{2}(\bar{g}^{\mu \rho}\bar{g}^{\nu \sigma}+\bar{g}^{\mu \sigma}\bar{g}^{\nu \rho})-\frac{1}{3}\bar{g}^{\mu \nu}\bar{g}^{\rho \sigma},
\end{eqnarray}
where $\bar{g}^{\rho \sigma}=g^{\rho \sigma}-\frac{p^\rho p^\sigma}{p^2}$. 

\begin{figure}[htbp]
    \centering
    \includegraphics[width=1.0\linewidth]{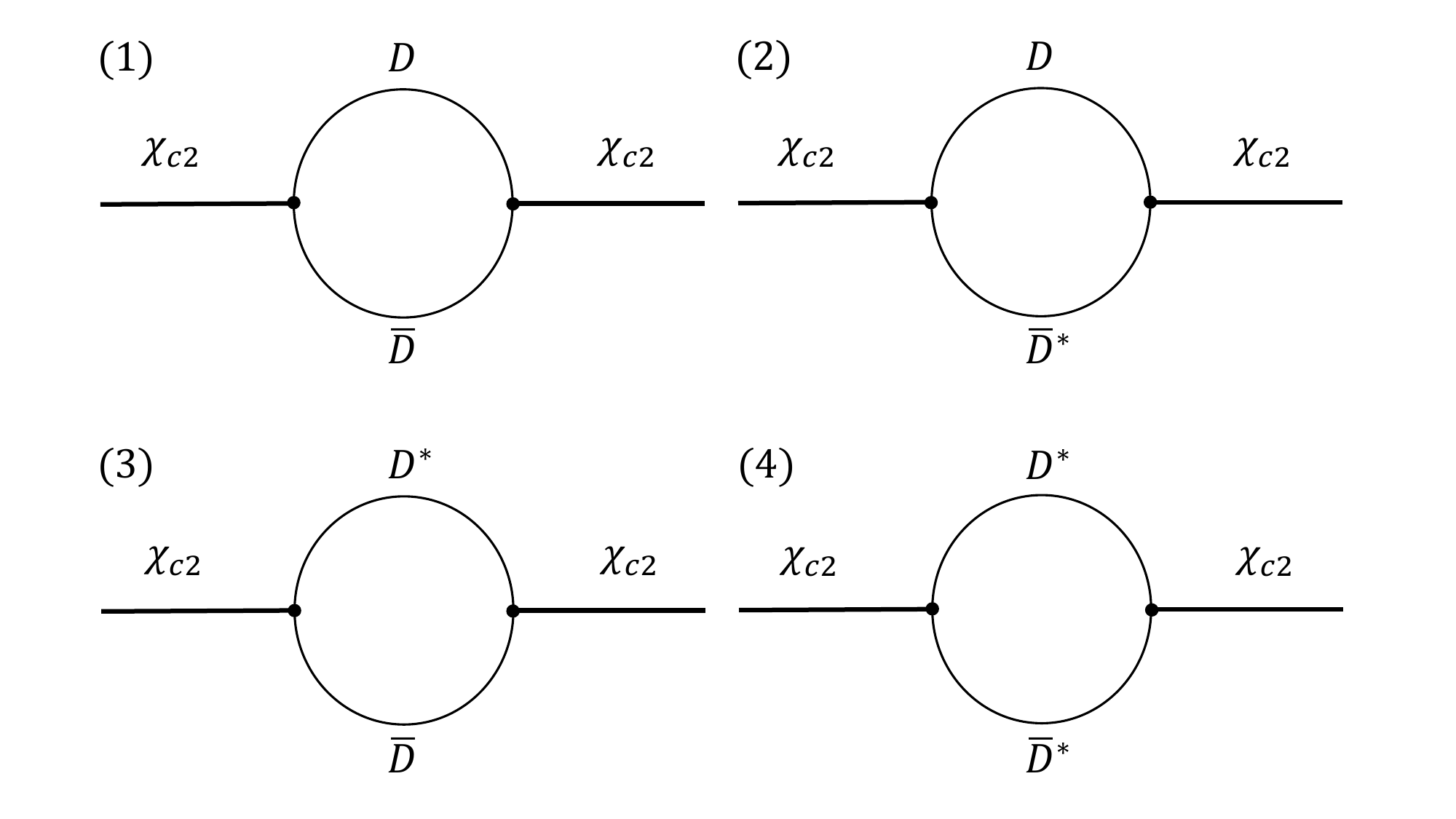}
    \caption{Schematic diagram showing the charmed-meson loop contributions to the $\chi_{c2}(1P)$ self-energy: (1) $D\bar{D}$, (2) $D\bar{D}^{*}$, (3) $D^{*}\bar{D}$, and (4) $D^{*}\bar{D}^{*}$.}
    \label{fig:chic2loop}
\end{figure}

The loop diagrams for the $\chi_{c2}(1P)$ meson are shown in Fig.~\ref{fig:chic2loop}. Using Eq.~(\ref{eq:projector2}), the contributions from $D\bar{D}$, $D^*\bar{D}$, $D\bar{D}^*$ and $D^*\bar{D}^*$ loops to the $\chi_{c2}(1P)$ self-energy are
\begin{eqnarray}
     \Pi_{\chi_{c2}D\bar{D}}(p^2)&=&\frac{2i}{5}P^{\text{mass}}_{\mu \nu \rho \sigma}g_{\chi_{c2}D\bar{D}}^2 \notag\\
        &&\times\int \frac{d^4 q}{(2 \pi)^4} \frac{(q-p)^{\mu} q^{\nu} q^{\rho} (q-p)^{\sigma}}{(q^2-m_D^2)((q-p)^2-m_{\bar{D}}^2)}\notag\\
        &&\times \mathcal{F}_{D}\big(q^2,m_{D}^{2}\big) \mathcal{F}_{\bar{D}}\big((p-q)^2,m_{\bar{D}}^{2}\big),\label{eq:chic2dd}
        \end{eqnarray}
        \begin{eqnarray}
        \Pi_{\chi_{c2}D^*\bar{D}}(p^2)&=&\Pi_{\chi_{c2}D\bar{D}^{*}}(p^2)\notag\\
        &=&\frac{2i}{5}P^{\text{mass}}_{\rho \sigma \lambda \kappa}\cdot 2 g_{\chi_{c2}D\bar{D}^{*}}^2\varepsilon_{\mu
        \nu \alpha \beta}\varepsilon_{\mu^{'}\nu^{'}\alpha^{'}\beta^{'}}p^{\alpha}p^{\alpha^{'}}\notag\\
        &&\times \int \frac{d^4 q}{(2 \pi)^4} \frac{(q-p)^{\beta} q^{\sigma} q^{\kappa} (q-p)^{\beta^{'}}\tilde{g}^{\nu \nu^{'}}(q)}{(q^2-m_{D^*}^2)((q-p)^2-m_{\bar{D}}^2)}g^{\rho\mu}g^{\lambda \mu^{'}}\notag\\
        &&\times \mathcal{F}_{D^{*}}\big(q^2,m_{D^{*}}^{2}\big) \mathcal{F}_{\bar{D}}\big((p-q)^2,m_{\bar{D}}^{2}\big),\label{eq:chic2dds}\\
        \Pi_{\chi_{c2}D^*\bar{D}^*}(p^2)&=&\frac{2i}{5}P^{\text{mass}}_{\mu \nu \rho \sigma }\cdot 2 g_{\chi_{c2}D^{*}\bar{D}^{*}}^2 \notag\\
        &&\times \int \frac{d^4 q}{(2 \pi)^4}\frac{\tilde{g}^{\mu \sigma}(q)\tilde{g}^{\nu \rho}(p-q)+\tilde{g}^{\mu \rho}(q)\tilde{g}^{\nu \sigma}(p-q)}{(q^2-m_{\bar{D}^*}^2)((p-q)^2-m_{D^*}^2)}\notag\\
        &&\times \mathcal{F}_{D^{*}}\big(q^2,m_{D^{*}}^{2}\big) \mathcal{F}_{\bar{D}^{*}}\big((p-q)^2,m_{\bar{D}^{*}}^{2}\big),\label{eq:chic2dsds}
\end{eqnarray}
and the total contribution to $\chi_{c2}(1P)$ is
\begin{eqnarray}
     \label{eq:chic2tot}
     \Pi^{\text{total}}_{\chi_{c2}}=\Pi_{\chi_{c2}D\bar{D}}+\Pi_{\chi_{c2}D^*\bar{D}}+\Pi_{\chi_{c2}D\bar{D}^{*}}+\Pi_{\chi_{c2}D^*\bar{D}^*}.
\end{eqnarray}

Before evaluating the in-medium mass $M^{*}_{\chi_{cJ}}$, the free space mass $M_{\chi_{cJ}}$ and the real part of $\Pi_{l}(p^2)$ in Eq.~(\ref{eq:scalarselfenergy}) are used to determine the corresponding bare mass:
\begin{eqnarray}\label{eq:baremass}
	\big(M_{\chi_{cJ}}^{0}\big)^{2}= (M_{\chi_{cJ}})^{2} - \sum_{l} \Re \Pi_{l}\big[(M_{\chi_{cJ}})^{2}\big],
\end{eqnarray}
where we neglect the possible widths of $\chi_{cJ}(1P)$ and $D^{(*)}$ mesons when treating them as vacuum inputs.

We note here that Eq.~(\ref{eq:scalarselfenergy}) is the key ingredient of the unquenched loop 
calculation. On the one hand, when combined with the physical masses of the $1P$-wave charmonia, those integrals in free space yield the corresponding bare masses, which are assumed to remain unaffected by medium modifications due to their purely heavy-quark composition. On the other hand, as the $D^{(*)}$ meson mass is modified in the QMC model, these self-energy contributions naturally acquire medium modifications, leading to the altered mass as indicated in Eq.~(\ref{eq:massinmedium}).

\section{Numerical Results}\label{sec:results}

In this section, we present a detailed analysis of the unquenched loop integration. The mass shifts of $\chi_{cJ}(1P)$ are then evaluated at finite nuclear density through these loops, with emphasis on the effects of different intermediate channels.

To implement the loop integration in Eq.~(\ref{eq:scalarselfenergy}), we first specify the physical masses and coupling constants as numerical inputs. The free-space masses of the $\chi_{c0}(1P)$,~$\chi_{c1}(1P)$ and $\chi_{c2}(1P)$ mesons are taken from the Particle Data Group~\cite{ParticleDataGroup:2024cfk} as $3414.71$ MeV, $3510.67$ MeV and $3556.17$ MeV, respectively. Another essential input is the coupling constant for the interaction between $\chi_{cJ}(1P)$ and $D^{(*)}\bar{D}^{(*)}$ in Eq.~(\ref{eqs:chicjDD}), which are related to the gauge coupling $g_{P}$ via~\cite{Chen:2010re,Huang:2021kfm}:
\begin{eqnarray}
  g_{\chi_{c0} DD}&=&2\sqrt{3} g_{P}\sqrt{m_{\chi_{c0}}} m_{D}, \,
  g_{\chi_{c0} D^{*} D^{*} } =\frac{2}{\sqrt{3}} g_{P} \sqrt{m_{\chi_{c0}}} m_{D^{*}},\notag\\
  g_{\chi_{c1} D D^{*}} &=& 2\sqrt{2} g_{P} \sqrt{m_{\chi_{c1}} m_{D} m_{D^{*}}}, \,
  g_{\chi_{c2} D D} =2g_{P} \frac{\sqrt{m_{\chi_{c2}}}}{m_{D}}, \notag\\
  g_{\chi_{c2} D D^{*}} &=& g_{P} \sqrt{\frac{m_{\chi_{c2}}}{m_{D^{*}}^3 m_{D}}},\,
  g_{\chi_{c2} D^{*} D^{*}}=4g_{P} \sqrt{m_{\chi_{c2}}} m_{D^{*}},
\end{eqnarray}
where the gauge coupling is given by $g_{P}=-\sqrt{\frac{m_{\chi_{c0}}}{3}}\frac{1}{f_{\chi_{c0}}}$ with $f_{\chi_{c0}}=0.51$ GeV, as obtained by QCD sum rule analysis \cite{Colangelo:2002mj,Colangelo:2003sa}. 

A key element is the form factor $\mathcal{F}_{i}(q^2,m_E^2)=\left(\frac{m_E^2-\Lambda^2}{q^2-\Lambda^2}\right)^2$, which enters the self-energy integral. Here, the cutoff momentum is parameterized as $\Lambda = m_{E} + \alpha \Lambda_{\text{QCD}}$, where $\Lambda_{\text{QCD}}=220~\text{MeV}$ and $\alpha$ is a dimensionless parameter expected to be of order unity~\cite{Cheng:2004ru}. In this work, we treat $\alpha$ as a free parameter and vary it from $2$ to $4$ to examine its influence on the results, following the approach of Refs.~\cite{Zeminiani:2020aho,Qian:2023taw,Gao:2024qth}. We note that our choice of a dipole form factor slightly differs from that used in Refs.~\cite{Krein:2010vp,Tsushima:2011kh,Cobos-Martinez:2017vtr,Cobos-Martinez:2017woo,Cobos-Martinez:2020ynh,Zeminiani:2020aho,Zeminiani:2023gqc,Zeminiani:2024dyo}, where the form factor was introduced after performing the $q^{0}$ integration. As seen in Eqs.~(\ref{eq:chic0dsds}) and (\ref{eq:chic2dsds}), the present choice offers an alternative prescription for exploring form-factor effects, as discussed in Ref.~\cite{Zeminiani:2020aho}.

With coupling constants and the form factor specified above, the bare mass is evaluated using Eq.~(\ref{eq:baremass}). The resulting bare masses $M^{0}_{\chi_{cJ}}$ depend on the total spin $J$, the cutoff parameter $\alpha$, and the specific loop contributions included in the calculation. They are determined so as to reproduce the observed free-space masses of the corresponding charmonium states. A systematic summary of the cutoff dependence of the bare masses for different loop contributions and total spins is provided in Table~\ref{tab:baremass}.

\begin{table}[htbp]
\begin{center}
\caption{\label{tab:baremass}
Bare masses \( M^0_{\chi_{cJ}} \) (in MeV) determined using Eq.~(\ref{eq:baremass}) for different spin states and intermediate meson-loop contributions. Results are shown for several typical values of the cutoff parameter \(\alpha\). “Total” refers to the combined contribution from all relevant loops. Gray cells indicate total values.}
\renewcommand\tabcolsep{0.34cm}
\renewcommand{\arraystretch}{1.5}  % Adds some vertical padding
\begin{tabular}{
    c
    c
    >{\columncolor{white}}c
    >{\columncolor{white}}c
    >{\columncolor{white}}c}
\toprule
\toprule
 & \textbf{Hadronic loop} & \(\boldsymbol{\alpha = 2}\) & \(\boldsymbol{\alpha = 3}\) & \(\boldsymbol{\alpha = 4}\) \\
\hline
\multirow{3}{*}{\( M^0_{\chi_{c0}} \)} 
    & \( D\bar{D} \)         & 3467.2 & 3525.6 & 3591.1 \\
    & \( D^*\bar{D}^* \)     & 3427.6 & 3450.9 & 3488.1 \\
    & \cellcolor{gray!10}Total  & \cellcolor{gray!10}3481.2 & \cellcolor{gray!10}3564.8 & \cellcolor{gray!10}3670.6 \\
\hline
\multirow{2}{*}{\( M^0_{\chi_{c1}} \)} 
    & \( D\bar{D}^{*}~\text{or}~D^{*}\bar{D} \)     & 3558.4 & 3610.0 & 3664.3 \\
    & \cellcolor{gray!10}Total  & \cellcolor{gray!10}3605.6 & \cellcolor{gray!10}3706.7 & \cellcolor{gray!10}3811.7 \\
\hline
\multirow{4}{*}{\( M^0_{\chi_{c2}} \)} 
    & \( D\bar{D} \)         & 3556.7 & 3557.8 & 3560.0 \\
    & \( D^*\bar{D}~\text{or}~D\bar{D}^* \)       & 3557.0 & 3559.2 & 3563.5 \\
    & \( D^*\bar{D}^* \)     & 3771.8 & 4067.6 & 4432.9 \\
    & \cellcolor{gray!10}Total  & \cellcolor{gray!10}3861.6 & \cellcolor{gray!10}4188.1 & \cellcolor{gray!10}4574.4 \\
\bottomrule
\bottomrule
\end{tabular}
\end{center}
\end{table}

In nuclear matter, the effective mass of the charmed meson is first obtained using the QMC model. Its value at $\rho_0$ is given in Table~\ref{tab:bagparam} as an example, and its density dependence can be found in Ref.~\cite{Tsushima:1998ru}. For each hadronic loop—corresponding to different bare masses listed in Table~\ref{tab:baremass}—the integration in Eq.~(\ref{eq:scalarselfenergy}) is performed using the in-medium mass of the $D^{(*)}$ meson. In this case, $M^{*}_{\chi_{cJ}}$ is determined self-consistently through Eq.~(\ref{eq:massinmedium}), rather than by substituting the free-space value.

To quantify the medium effect, we define the mass shift $\Delta M$ as:
\begin{eqnarray}
     \Delta M_{\chi_{cJ}}=M^{*}_{\chi_{cJ}}-M_{\chi_{cJ}}.
\end{eqnarray}

\begin{figure}[htbp]
    \centering
    \includegraphics[width=1.0\linewidth]{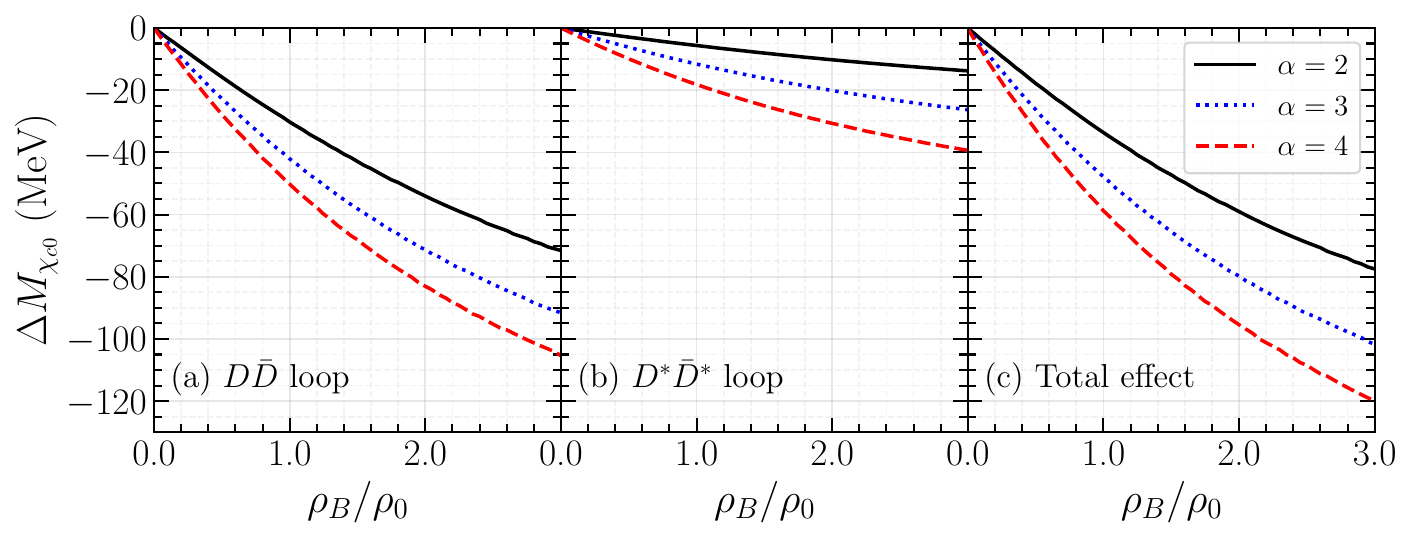}
    \caption{Mass shift of the $\chi_{c0}(1P)$ meson as a function of baryon density $\rho_{B}$ (in units of the normal nuclear matter density, $\rho_{0} = 0.15\,\mathrm{fm}^{-3}$) arising from the $D\bar{D}$ loop (a), the $D^{*}\bar{D}^{*}$ loop (b), and their combined effect (c); the solid, dotted, and dashed lines represent cutoff values $\alpha = 2$, $3$, and $4$, respectively.}
    \label{fig:chic0massshift}
\end{figure}

We now present the results for the mass shifts. For the $\chi_{c0}(1P)$ meson, we consider two intermediate loops: $D\bar{D}$ and $D^{*}\bar{D}^{*}$, as well as their combined effect, shown in Fig.~\ref{fig:chic0massshift}. At normal nuclear matter density $\rho_{0}$, the mass shift from the $D\bar{D}$ loop alone ranges from $-30.2$ MeV to $-50.2$ MeV, which is larger than that from the $D^{*}D^{*}$ channel alone ($-5.6$ MeV to $-18.2$ MeV) for different cutoff parameters. The total contribution from both channels, computed using Eq.~(\ref{eq:chic0totloop}), varies from $-33.6$ MeV to $-58.8$ MeV. In contrast to previous studies of in-medium $J/\psi$ and $\Upsilon$, where the vector-vector channel was found to yield an unexpectedly large contribution, we find that the $D^{*}\bar{D}^{*}$ loop contribution for $\chi_{c0}(1P)$ is less pronounced than that from the $D\bar{D}$ loop.

\begin{figure}[htbp]\centering
    \includegraphics[width=1.0\linewidth]{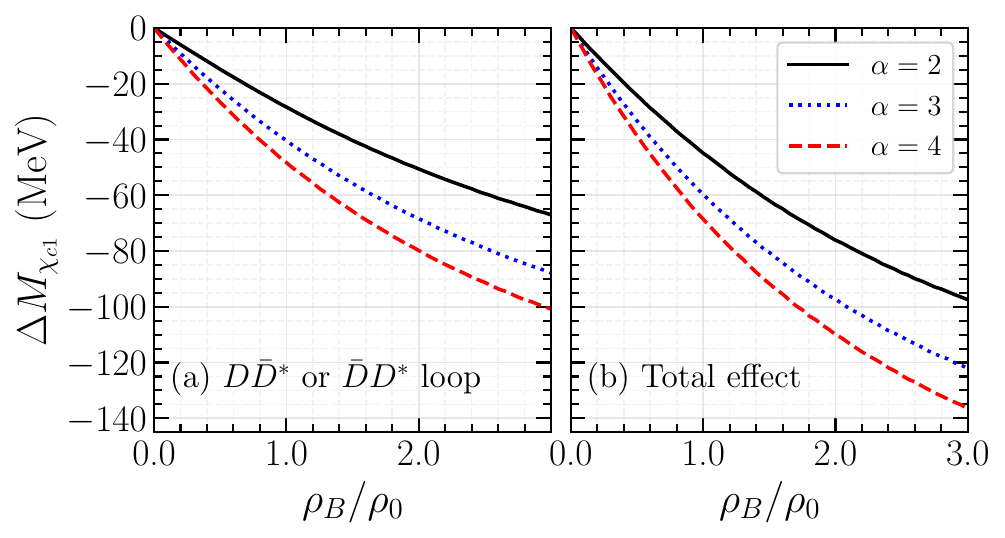}
  \caption{Mass shift of the $\chi_{c1}(1P)$ meson as a function of baryon density $\rho_{B}$ (in units of $\rho_{0}$) arising from the $D\bar{D}^{*}$ or the $D^{*}\bar{D}$ loop (a), and their combined effect (b); the solid, dotted, and dashed lines represent cutoff values $\alpha = 2$, $3$, and $4$, respectively.}
\label{fig:chic1massshift}
\end{figure}

For the $\chi_{c1}(1P)$ meson mass shift, the $D\bar{D}^{*}$ and $D^{*}\bar{D}$ channels corresponding to Eq.~(\ref{eq:chic1oneloop}), are evaluated at finite density and receive identical medium modifications for each individual loop. At $\rho_{0}$, the mass shift from either the $D\bar{D}^{*}$ or $D^{*}\bar{D}$ loop alone ranges from $-28.3$ MeV to $-48.3$ MeV, while their total contribution from Eq.~(\ref{eq:chic1totloop}) is between $-44.9$ MeV and $-68.6$ MeV. Figure~\ref{fig:chic1massshift} shows the separate contribution from either the $D\bar{D}^{*}$ or $D^{*}\bar{D}$ loop, as well as their sum, to the $\chi_{c1}(1P)$ mass shift. We note that the mass shift for the $\chi_{c1}(1P)$ meson is more significant than that for $\chi_{c0}(1P)$ in the region $\rho_{B} < 3\rho_{0}$.

\begin{figure}[htbp]
  \centering
\includegraphics[width=1.0\linewidth]{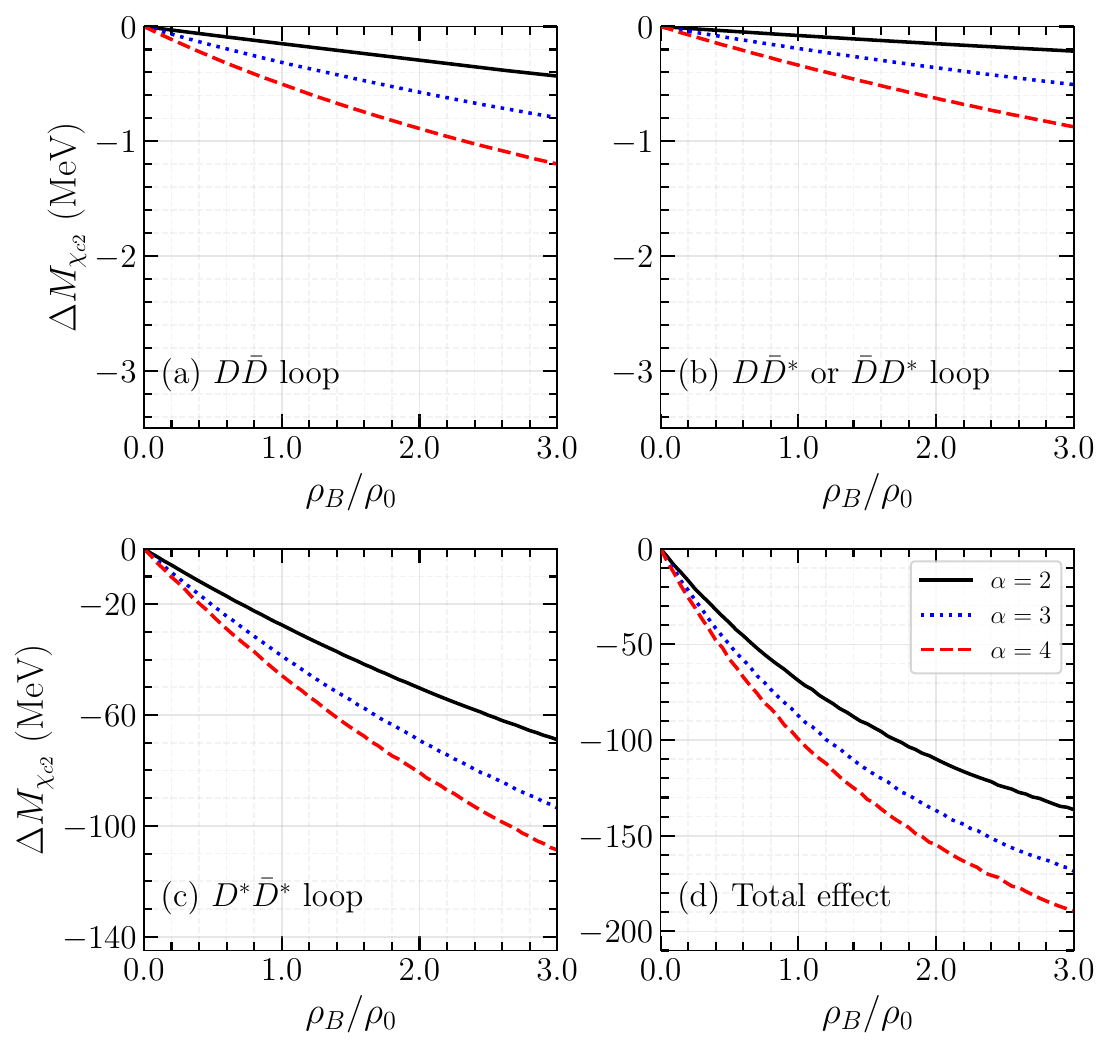}
\caption{Mass shift of the $\chi_{c2}(1P)$ meson as a function of baryon density $\rho_{B}$ (in units of $\rho_{0}$) due to (a) the $D\bar{D}$ loop, (b) the $D\bar{D}^{*}$ or the $D^{*}\bar{D}$ loop, (c) the $D^{*}\bar{D}^{*}$ loop, and (d) their total effect. The solid, dotted, and dashed lines represent cutoff values $\alpha = 2$, $3$, and $4$, respectively.}
  \label{fig:chi2massshift}
\end{figure}

The in-medium mass of the $\chi_{c2}(1P)$ meson in symmetric nuclear matter is obtained using Eqs.~(\ref{eq:chic2dd})-(\ref{eq:chic2tot}), corresponding to the contributions from the $D\bar{D}$, $D\bar{D}^{*}$ or $D^{*}\bar{D}$, $D^{*}\bar{D}^{*}$ loops, and their sum. As shown in Fig.~\ref{fig:chi2massshift}~(a)-(b), the mass shift of  $\chi_{c2}(1P)$ is negligible when considering only the $D\bar{D}$, $D\bar{D}^{*}$, or $D^{*}\bar{D}$ loops. However, Fig.~\ref{fig:chi2massshift}~(c) indicates that the dominant medium effect comes from the $D^{*}\bar{D}^{*}$ loop, ranging from $-27.4~\text{MeV}$ to $-45.7~\text{MeV}$ at $\rho_{0}$. When all four loops are included, the total mass shift of the $\chi_{c2}$ meson is between $-63.9~\text{MeV}$ and $-97.4~\text{MeV}$ at $\rho_{0}$, and its medium effect, shown in Fig.~\ref{fig:chi2massshift}~(d), is larger than that of $\chi_{c0}(1P)$ or $\chi_{c1}(1P)$ for $\rho_{B}<3\rho_{0}$. The contribution from the $D\bar{D}$, $D\bar{D}^{*}$ or $D^{*}\bar{D}$ loops alone is less significant for $\chi_{c0}(1P)$ or $\chi_{c1}(1P)$, which warrants further explanation. For the $D\bar{D}$ loop, this is due to its $D$-wave coupling to the $\chi_{c2}(1P)$ meson, whereas it couples via $S$-wave to $\chi_{c0}(1P)$. The $S$-wave coupling between the $D\bar{D}^{*}$/$D^{*}\bar{D}$ loop and the $\chi_{c1}(1P)$ is stronger than that for the $D$-wave $\chi_{c2}(1P) D\bar{D}^{*}$ coupling.

Several comments on the numerical results are in order. First, our calculations show that the in-medium mass modification of $D^{(*)}$ and $\bar{D}^{(*)}$ mesons leads to a net downward shift in the mass of the $1P$-wave charmonium triplet $\chi_{cJ}(1P)$, where the $D^{*}\bar{D}^{*}$ loop diagram is crucial for evaluating the $\chi_{c2}(1P)$ meson in nuclear matter. Thus, within the current framework, it is essential to include the vector-vector channel to properly account for the mass shift, in contrast to arguments for excluding them and retaining only the “lowest-order graphs” as discussed in Refs.~\cite{Krein:2017usp,Zeminiani:2020aho,Zeminiani:2023gqc}. Second, we present separate contributions from different loops and their corresponding bare masses—for example, the $\chi_{c0}(1P)$-$D\bar{D}$ loop in Eq.~(\ref{eq:chic0dd}) with $M_{\chi_{c0}}^{0}=3467.2~\mathrm{MeV}$ for $\alpha=2$—to facilitate direct comparison with other works using similar methods, such as Refs.~\cite{Zeminiani:2020aho,Zeminiani:2023gqc}. However, these results should not be confused with the partial contribution of a specific loop in the total combined effect.

\begin{figure}[htbp]\centering
\includegraphics[width=1.0\linewidth]{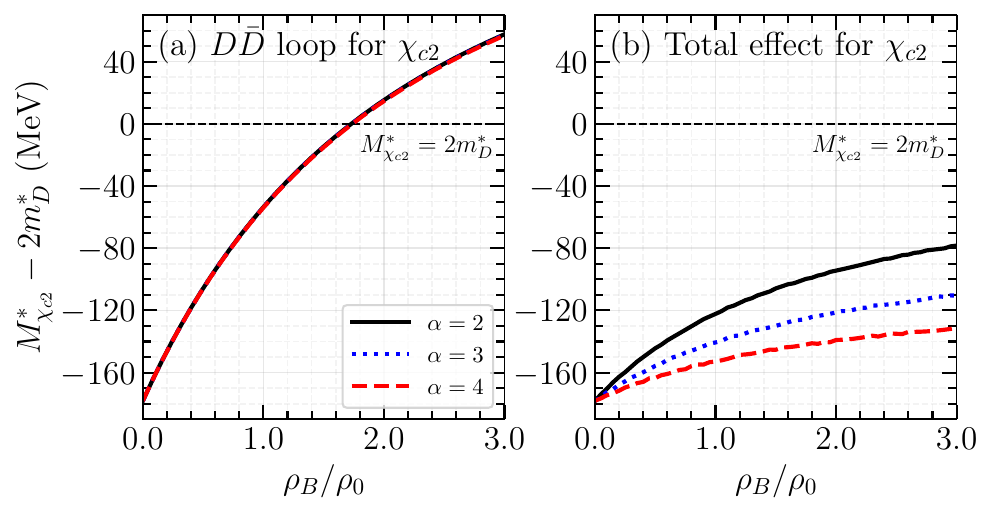}
  \caption{The mass difference of $\chi_{c2}(1P)$ meson and the effective mass threshold of $D\bar{D}$ as a function of baryon density $\rho_{B}$ (in units of $\rho_{0}$), with contribution from only the $D\bar{D}$ loop (left panel) and the total contribution (right panel). The solid, dotted, and dashed lines correspond to cutoff parameters $\alpha=2$, $3$, and $4$, respectively.}
\label{fig:realcondition}
\end{figure}

Finally, similar to the sequential suppression scenario in which feed-down from $\psi(2S)$ and $\chi_{cJ}(1P)$ to $J/\psi$ disappears at finite temperature, level crossing with increasing nuclear matter density was discussed in Ref.~\cite{Hayashigaki:2000es}. There, it was suggested that the $\chi_{c2}(1P)$ mass would cross the in-medium $D\bar{D}$ mass threshold at approximately $2\rho_{0}$ if the free-space charmonium mass were used. This argument was examined from the perspective of the partial decay widths of excited charmonium states into $D\bar{D}$, revealing a non-trivial density dependence~\cite{Friman:2002fs}. In the present work, we go a step further by including the $\chi_{c2}(1P)$ mass shift. If one were to exclude contributions from loops containing a $D^{*}$ meson, the effective mass of $\chi_{c2}(1P)$ would exceed the $D\bar{D}$ threshold in the medium for $\rho_{B} > 1.8\rho_{0}$, as shown in Fig.~\ref{fig:realcondition} (a). However, the mass shift of $\chi_{c2}$ significantly alters this picture when the total effect is included (Fig.~\ref{fig:realcondition} (b)), preventing level crossing in the density region $0$–$3\rho_{0}$ covered by our calculation.

\section{Summary}
\label{sec:summary}

The medium modification of charmonia provides a key theoretical input for predicting $(c\bar{c})$-nucleus bound states and interpreting the observed suppression of $J/\psi$ production. Previous studies of in-medium charmonia within the QMC model have incorporated an unquenched picture, yet contributions from vector-vector loops were omitted due to their unexpectedly large influence on the mass shift~\cite{Krein:2017usp,Zeminiani:2020aho,Zeminiani:2023gqc}. However, this approach—which retains only the “lowest-order graphs”—may be overly restrictive, especially in the case of the $P$-wave charmonium state $\chi_{c2}(1P)$, whose coupling to the vector-vector channel is pronounced.

In this work, the in-medium mass shifts of the $1P$-wave charmonium states $\chi_{cJ}(1P)$ in cold symmetric nuclear matter is studied using the QMC model and heavy quark effective field theory. Medium modifications are incorporated through density-dependent masses of $D^{(*)}$ and $\bar{D}^{(*)}$ mesons obtained from the QMC model. Within the unquenched picture, the self-energies of the $\chi_{cJ}(1P)$ states receive contributions from various hadronic loops, including $D\bar{D}$, $D^*\bar{D}$, $D\bar{D}^*$, and $D^*\bar{D}^*$. Notably, the vector-vector channel—previously acknowledged but discarded—is found to yield a substantial mass shift for $\chi_{c0}(1P)$ and $\chi_{c2}(1P)$. Our results indicate that such state-dependent mass shifts may help clarify the stepwise suppression scenario at finite density.

Phenomenologically, the in-medium $D\bar{D}$ mass threshold was expected in Ref.~\cite{Hayashigaki:2000es} to sequentially drop below the effective masses of $\psi(2S)$, $\chi_{c2}(1P)$, $\chi_{c1}(1P)$, and $\chi_{c0}(1P)$ as density increases. In contrast to this stepwise suppression picture, our results—which directly incorporate $P$-wave charmonium mass shifts—reveal that the $D\bar{D}$ threshold remains above the mass spectrum of $P$-wave charmonia when the contribution from vector-vector loops such as $D^{*}\bar{D}^{*}$ (particularly relevant for $\chi_{c2}(1P)$) is taken into account.

Needless to say, finite temperature effects also play an important role in heavy-ion collisions. In the future, we plan to extend this study to explore medium effects at finite temperature using the QMC model, as well as unquenched effects for other charmonium states. On the experimental side, yield estimates for future experiments suggest that $\chi_{c0}(1P)$ and $\chi_{c2}(1P)$ could exhibit even higher daily event rates than $J/\psi$ and $\eta_c$~\cite{Lee:2003hm,Morita:2007hv}, making their in-medium mass shifts more accessible in experiments such as the Compressed Baryonic Matter experiment at FAIR-GSI. Moreover, probing the properties of the quark-gluon plasma at high density is a key objective of the Beam Energy Scan program at RHIC~\cite{Bzdak:2019pkr}. Our results may also provide a theoretical input for modeling various cold nuclear matter effects when both temperature and density dependencies become significant~\cite{Zhao:2022ggw}.

\begin{acknowledgments}

Ze-Hua Zhang would like to thank Dian-Yong Chen, Jin-Niu Hu, Feng Li, Zhan-Wei Liu, Si-Qiang Luo, Ri-Qing Qian, Kai-Jia Sun and Kazuo Tsushima for useful discussions. This work is supported by the National Natural Science Foundation of China under Grant Nos. 12335001, 12247101, and 12405098,  the ‘111 Center’ under Grant No. B20063, the Natural Science Foundation of Gansu Province (No. 22JR5RA389, No. 25JRRA799), the Talent Scientific Fund of Lanzhou University, the fundamental Research Funds for the Central Universities, the project for top-notch innovative talents of Gansu province, and Lanzhou City High-Level Talent Funding.
\end{acknowledgments}

\bibliography{ref1.bib}

\end{document}